\documentclass[
reprint,  %%% ARXIV
 amsmath,amssymb,
 aps,
 prxquantum,
]{revtex4-2}
\usepackage[utf8]{inputenc}
\usepackage[version=3]{mhchem}
\usepackage{amsmath}
\usepackage{color}
\usepackage{titlesec}
\usepackage{soul}
\usepackage{braket}
\usepackage{xr-hyper}
\usepackage{hyperref}
\usepackage{graphicx}% Include figure files
\usepackage{dcolumn}% Align table columns on decimal point
\usepackage{bm}% bold math
\usepackage{tikz}
\usepackage{ragged2e}
\usepackage{kantlipsum}
 \usepackage{setspace}
 \usepackage{upgreek}
\usepackage{capt-of}
\usepackage{float}
\usepackage{cancel}
\usepackage[normalem]{ulem}

\allowdisplaybreaks % Allow automatic breaks for equations

\usetikzlibrary{matrix}
\usepackage{eqparbox}
\newbox\matrixcellbox
\tikzset{center align per column/.style={nodes={execute at begin
            node={\setbox\matrixcellbox=\hbox\bgroup},
            execute at end
            node={\egroup\eqmakebox[\tikzmatrixname\the\pgfmatrixcurrentcolumn][c]{\copy\matrixcellbox}}}},
}

\titleformat{\section}{\centering\normalfont\bfseries}{\thesection}{1em}{}
\titlespacing{\section}{0pt}{*2}{*1}
\titlespacing{\subsection}{0pt}{*2}{*1}

\usepackage{xcolor}
\usepackage[draft,inline,nomargin,index]{fixme}

\fxsetup{theme=color,mode=multiuser}

\FXRegisterAuthor{ofer}{aOfer}{\color{brown} Ofer} 
\FXRegisterAuthor{bankim}{abankim}{\color{red}  Bankim}
\FXRegisterAuthor{sasha}{asasha}{\color{orange} \bf Sasha}

\definecolor{cadmiumgreen}{rgb}{0.0, 0.68, 0.24}

\newcommand{\OD}{\mathrm{OD}}

\newcommand {\rmi}{{i}}
\newcommand {\e}{{e}}
\newcommand {\rmd}{{d}}

\newcommand{\xo}{x_\mathrm{o}}

\newcommand{\zero}{0}
\newcommand{\len}{L}

\newcommand{\omegd}{\nu}
\begin{document}
\onecolumngrid
\twocolumngrid
\title{Multiband dispersion and warped vortices of strongly interacting photons}

\author{Bankim Chandra Das}
\author{Dmytro Kiselov}
\author{Lee Drori}
\author{Ariel Nakav}
\author{Alexander Poddubny}
\author{Ofer Firstenberg}
\affiliation{Department of Physics of Complex Systems, Weizmann Institute of Science, Rehovot 7610001, Israel}  
%\date{\today}
\begin{abstract}
We present a theoretical study of quantum correlations between interacting photons realized through co-propagating Rydberg polaritons. We show that the spatial evolution of the $n$-photon wavefunction is governed by a multiband dispersion featuring one massive mode and multiple massless modes with degenerate Dirac points and $n$-fold rotational symmetry.  The resulting band structure is warped, departing from the single-band, parabolic approximation commonly assumed for interacting polaritons. 
Our analytical results are supported by rigorous numerical modeling that fully accounts for photon propagation inside the finite atomic medium. 
These findings advance the understanding of multi-photon interactions and support the development of future multi-photon control tools.  
\end{abstract}

\maketitle

Rydberg polaritons, formed by coupling propagating photons to Rydberg atoms in a dense atomic ensemble, provide a powerful platform for studying quantum interactions between a few particles \cite{Murray2016Jan,Firstenberg2016Jun}. Their strong interactions arise from the long-range dipolar coupling between their Rydberg constituents. In the quantum nonlinear optics regime \cite{Chang2014Sep}, Rydberg polaritons have been used to suppress the transmission of more than one propagating photons \cite{Peyronel2012,Hofferberth2014SinglePhotTransPRL,Rempe2014SinglePhotSwitchPRL,Ornelas2021} and to demonstrate large conditional phase shifts \cite{Tiarks2016SciAdv,Thompson2017Nature}, bound states~\cite{firstenberg2013attractive,Hofferberth2018,Liang2018}, and multimode interactions \cite{Busche2017Jul,Clark2019Jul} of two or three polaritons. Recently, stronger interactions enabled the observation of vortex-antivortex pairs in the two-polariton wavefunction $\psi(x_1,x_2)$ and vortex rings and tubes in the three-polariton wavefunction $\psi(x_1,x_2,x_3)$~\cite{DroriScience2023}. Here, $x_j$ are the photon coordinates along the propagation direction, \textit{i.e.}, in one spatial dimension.

These developments establish Rydberg polaritons as a promising platform for multiphoton control and deterministic quantum logic. They also highlight the need for a rigorous theoretical framework to quantitatively describe quantum correlations arising from the propagation and interaction of several photons, including vortex formation. 
While previous theoretical models of interacting Rydberg polaritons --- such as the adiabatic potential approximation~\cite{Jachymski2016}, quantum field theory
and Green function approaches \cite{GullansPRL2016, Grankin2018,Kalinowski2024}, and waveguide quantum electrodynamics models \cite{CanevaChang2015NJP,Chen2020,sheremet2023,Chang2022} --- offer valuable insights, they do not fully capture the spatial evolution of few-polariton wavefunctions in an extended medium.
Although the model in Ref.~\cite{firstenberg2013attractive}
successfully describes two-polariton dynamics, it neither predicts the three-photon vortices observed in Ref.~\cite{DroriScience2023} nor captures their unique symmetry reported here. Moreover, its extension to higher-order correlations remains unclear.

Here, we study the stationary wavefunction  $\psi(x_1,...,x_n)$ of interacting polaritons in a spatially extended medium. We reveal that $\psi$ follows a multiband Dirac-like dispersion in spatial momentum space, describing the dependence of the center-of-mass wavevector $K(\kappa_1\ldots \kappa_n)$ on the relative wavevectors $\kappa_1\ldots \kappa_n$. In the minimal model capturing the full symmetry of the problem, the dispersion features one symmetric massive mode  and $n-1$ {gapped} antisymmetric modes. For $n\ge3$, the antisymmetric modes are point-degenerate at zero momentum and exhibit linear dispersion.

\begin{figure}[tb]
\centering\includegraphics[width=0.48\textwidth,trim={0cm 0cm 0cm 0cm},clip]{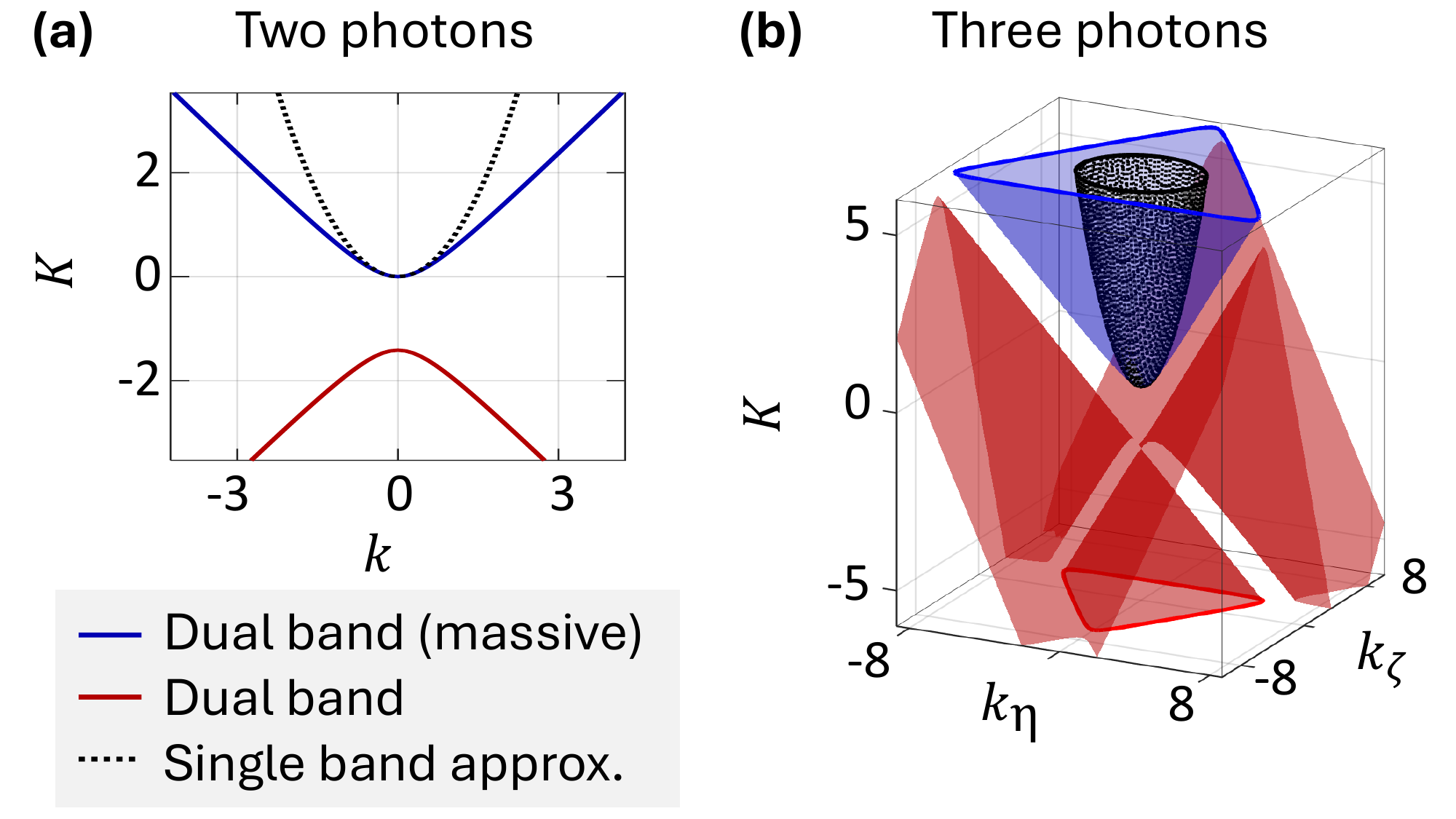}

\caption{%\textbf
{Analytical momentum-space dispersion.} 
%\textbf
{(a) Two photons:} Center-of-mass dispersion $K(k)$ for two non-interacting polaritons. The dashed line represents the single-band Schr\"odinger approximation of the massive polariton branch, while the solid lines show the full two-band Dirac-like dispersion. 
%\textbf
{(b) Three photons:} Center-of-mass dispersion $K(k_{\upeta},k_{\upzeta})$ for three polaritons, highlighting the emergence of trigonal warping. The black surface represents the single-band Schr\"odinger approximation, which fails to capture the warping. $K(k)$ and $K(k_{\upeta},k_{\upzeta})$ are found from  Eqs.~\eqref{eq:K2b} and \eqref{eq:K3},  and all units in $\rho g^2/(c\Delta)$.
}\label{fig:dispersion}
\end{figure}

As an approximation, these antisymmetric modes can be eliminated by assuming they adiabatically follow the massive mode, thereby reducing the model to a Schr\"odinger-like equation for a single massive mode in $n-1$ dimensions with an $(n!)$-fold symmetry. This approximation has already provided valuable insight into the polariton dynamics in terms of two-photon and three-photon bound states \cite{firstenberg2013attractive,Liang2018}. However, as shown in Figs.~\ref{fig:dispersion} and \ref{fig:Dirac}, the parabolic single-band approximation fails to retain key features of the multiband structure, particularly the linear-dispersion regions and the reduced $n$-fold discrete rotational symmetry. Both are essential for describing the evolution in finite media and capturing the effects of interaction-dependent group velocity.

The multiband dispersion significantly impacts photon correlations starting from $n=3$ photons. It reduces the wavefunction's rotational symmetry from $3!=6$ to 3, by distinguishing between cases where a photon pair propagates ahead of or behind a single photon. 
This symmetry reduction was previously detected at a few-percent level in photon correlation functions with a single atom or an effective superatom (a fully blockaded Rydberg ensemble) \cite{Hofferberth2018,Denisov2002,Koch2011}.
Here, we develop analytical and numerical models that capture these dynamics and show that this asymmetry can be large and manifest as a trigonal warping of the vortex ring formed by three-photon interactions. 

\section*{Results}
\subsection{Multiband model for two and three photons}
\noindent We begin with a minimal analytical model that correctly captures the symmetry of few-polariton states in the stationary regime.
Consider a medium of three-level ladder-type atoms hosting Rydberg polaritons under conditions of electromagnetically induced transparency \cite{Gorshkov2011Sep,Peyronel2012}. The  single-polariton propagation along the medium is governed by $H(x)\psi=\rmi \partial_t\psi$, where 
\begin{equation}\label{eq:H}
	H(x)= 	
	{-}\begin{pmatrix}
	ic\partial_x &  \sqrt{\rho}g & 0\\	
	 \sqrt{\rho}g   & \Delta+\rmi\Gamma &  \Omega \\
	0& \Omega & \delta+\rmi\gamma\\
		\end{pmatrix}
\end{equation}
is the effective Hamiltonian for the atoms and photon.
The polariton wavefunction $\psi(x)=[E(x),P(x),S(x)]^T$ consists of amplitudes $E(x)$ for a single photon (all atoms in the ground state $|\mathrm{g}\rangle$), $P(x)$ for one atom in the intermediate state $|\mathrm{p}\rangle$, and $S(x)$ for one atom in the Rydberg state $|\mathrm{s}\rangle$.
Here, $g$ is the coupling strength of the photon to the $|\mathrm{g}\rangle$$\leftrightarrow$$|\mathrm{p}\rangle$ transition, $\rho$ is the medium density, $\Omega$ is the Rabi frequency of the $|\mathrm{p}\rangle$$\leftrightarrow$$|\mathrm{s}\rangle$ driving, $\Gamma$ is the decay rate of $|\mathrm{p}\rangle$, $\gamma$ is the decoherence rate of $|\mathrm{s}\rangle$, and $\Delta$ and $\delta$ are the one-photon and two-photon detunings. We consider $|\Delta|\gg \Gamma$ and, for simplicity, assume $\Gamma=\delta=\gamma=0$.

Importantly, realistic experiments \cite{firstenberg2013attractive,Hofferberth2018,Liang2018,DroriScience2023} are conducted under the steady state conditions, $\partial_t\psi=0$, allowing us to simplify the model by eliminating the $|\mathrm{p}\rangle$ states from the equation $H(x)\psi=0$ via $P=-(\sqrt{\rho}g E+\Omega S)/\Delta$. The resulting Hamiltonian with eliminated $P$ states reads 
\begin{equation}\label{eq:HP}
	H(x)= 	
		\begin{pmatrix}
	\omegd-ic\partial_x &  \frac{\sqrt{\rho} g\Omega }{\Delta}\\	
\frac{\sqrt{\rho} g\Omega}{\Delta}   & \frac{\Omega^2}{\Delta}
		\end{pmatrix}
\end{equation}
for $[E(x),S(x)]^T$. Here,  the shift $\omegd={\rho g^2/\Delta}$ stems from the photon hybridization with the atomic states.

The Hamiltonian~\eqref{eq:HP} can be straightforwardly generalized to two and three interacting polaritons, characterized by multicomponent wavefunctions. For example, two polaritons are represented by four wavefunction components, $EE$, $ES$, $SE$, and $SS$, corresponding to each polariton residing either in the propagating-photon state or Rydberg state. The two-polariton Hamiltonian can be written as $\mathcal{H}(x_1,x_2)=H(x_1)\otimes I+I\otimes H(x_2)+\mathcal V$,
where the interaction potential $\mathcal V$ acting on $SS$ originates from van-der-Walls (vdW) interactions between Rydberg atoms and is given by Eq.~\eqref{eq:V}. 

In the reduced $E$-$S$ subspace, we further simplify the equations by expanding to lowest nonvanishing order in $\Omega^2$, valid for our parameters $\Omega/(\sqrt{\rho}g)\sim 10^{-3}\ll 1$. The 
coupling of the $EE$ component to $ES$ and $SE$ can then be neglected. The equations for the remaining $[ES,SE,SS]$ components read
\begin{equation}\label{eq:full2}
\begin{pmatrix}
  \omegd-\rmi c\partial_{x_{1}} & 0 & \frac{\sqrt{\rho}\, g \Omega}{\Delta} 
\\
  0 & \omegd-\rmi c\partial_{x_{2}} & \frac{\sqrt{\rho}\, g \Omega}{\Delta} 
\\
  \frac{\sqrt{\rho}\, g \Omega}{\Delta} & \frac{\sqrt{\rho}\, g \Omega}{\Delta} & \frac{2 \Omega^{2}}{\Delta}+\frac{ C_6}{|x_1-x_2|^6} 
\end{pmatrix}\begin{pmatrix}
ES\\SE\\SS
\end{pmatrix}=0\:.
\end{equation}

We can now eliminate the $SS$ state to obtain the equation governing the two-polariton propagation:
\begin{equation}\label{eq:two}
   {\rmi} c\begin{pmatrix}        
    {\partial_{x_1}}&0\\    
    0&{\partial_{x_2}}
    \end{pmatrix}
\psi{-}\omegd\psi{+}\frac{V_2}{2}\begin{pmatrix}
     1&1\\1&1
     \end{pmatrix}
       \psi=0\:,
\end{equation}
with  $\psi=[ES,SE]$, where the renormalized interaction potential $V_2$ is given below. 
Similarly, we find for three polaritons:
\begin{equation}\label{eq:three}
  {\rmi}c\begin{pmatrix}        
    {\partial_{x_1}}&0&0\\    
    0&{\partial_{x_2}}&0\\
    0&0&{\partial_{x_3}}\\
    \end{pmatrix}
\psi{-}\omegd\psi{+}\frac{V_3}{3}  \begin{pmatrix}
     1&1&1\\1&1&1\\1&1&1
     \end{pmatrix}     \psi=0,
\end{equation} 
with $\psi=[ESS,SES,SSE]$ corresponding to a single propagating photon with the other excitations in the Rydberg states.\

A key feature of Eqs.~\eqref{eq:two} and \eqref{eq:three} is the all-to-all interaction between the wavefunction components, arising from their coupling to the state with all excitations in the Rydberg state ($SS$ or $SSS$). The interaction potentials $V_2$ and $V_3$  depend on the polariton coordinates,
\begin{equation}\label{eq:Vn}
   {V_{n}=\omegd \biggl( 1+\frac{2}{n}\sum_{i,j=1; i\ne j}^n \frac{r_\mathrm{b}^6}{|x_i-x_j|^6} \biggl)^{-1}}
\end{equation}
for $n\ge 2$ polaritons. 
The potentials vanish inside the `Rydberg blockade' volume, when any two polaritons are closer than the blockade radius {$r_\mathrm{b}=(C_6\Delta/2\Omega^2)^{1/6}$},  where $C_6$ is the vdW coefficient, and approach the constant value $\omegd$ at large separations. 
Notably, for $n=2$, $V_n$ reduces (up to sign and offset) to the effective potential used in the approximated Schr\"odinger model.

\begin{figure*}[tb]
\centering\includegraphics[width=1\textwidth,trim={0cm 0cm 0cm 0cm},clip]{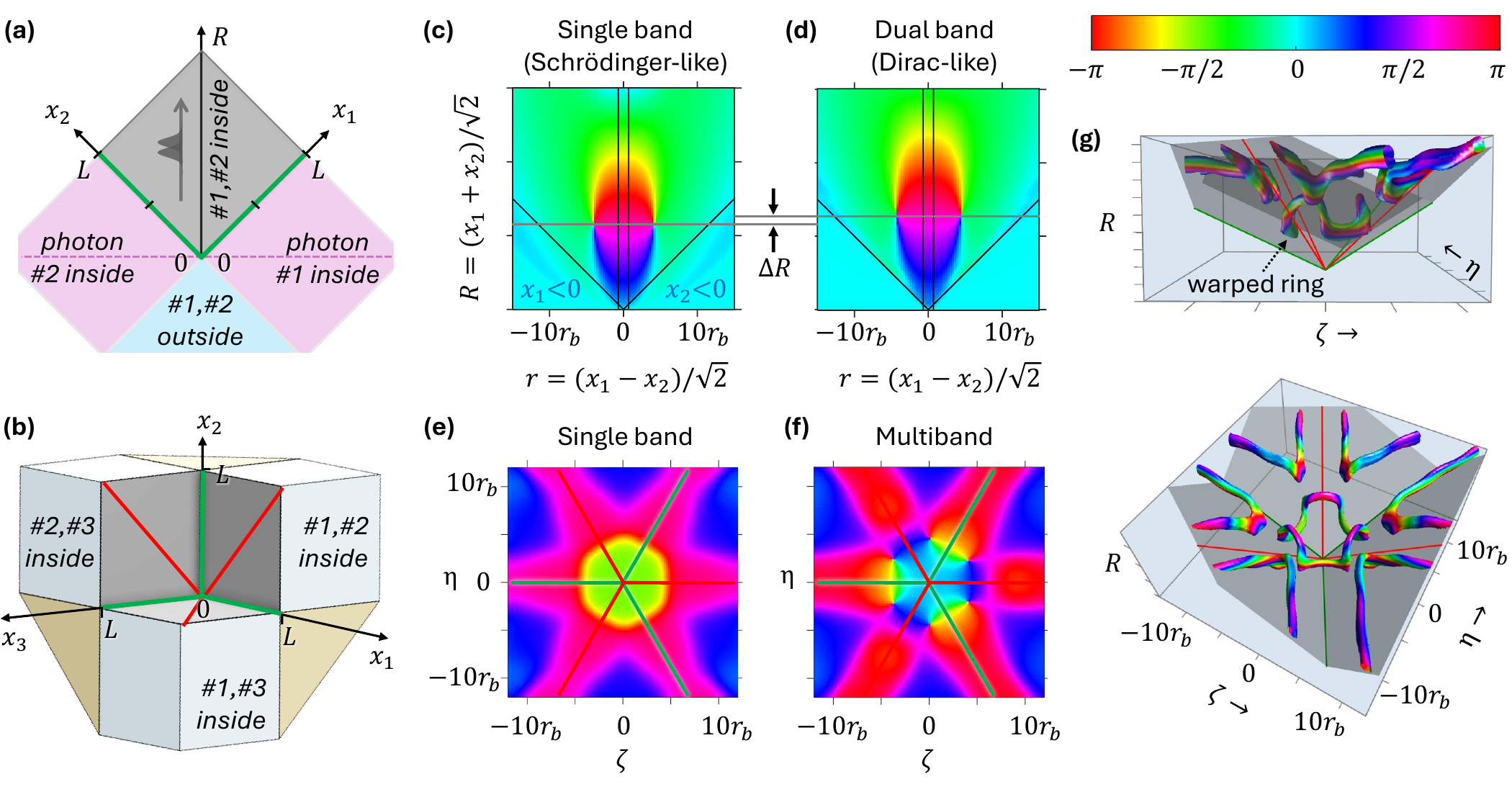}
\caption{%\textbf
{Analytical multiband model, real-space propagation.} 
%\textbf
{(a) Two-photon coordinate space:} Schematic showing the propagation center-of-mass coordinate $R$ and relative coordinate $r$. Two-photon interaction occurs only inside the medium ($\zero<x_{1,2}<\len$, gray area). The entry points $x_1=\zero$ and $x_2=\zero$ (green lines) serve as light-cone boundaries, below which two-photon correlations cannot develop.
%\textbf
{(b) Three-photon coordinate space:}  Interaction occurs already when photon pairs are inside the medium (light blue sections), while full three-photon interaction requires all $\zero<x_{1,2,3}<\len$. Regions above the green (red) lines correspond to a closely spaced photon pair propagating behind (ahead of) a single photon.
%\textbf
{(c,d) Two-photon phase profiles:} The phase of the two-photon wavefunction under the single-band model (c) shows a vortex-antivortex pair at finite propagation distance $R$ (gray line) but also unphysical correlations outside the light-cone boundaries ($x_{1,2}<\zero$).  The full two-band dispersion (d) corrects the unphysical behavior, confining correlations within the light cone and introducing a propagation delay $\Delta R$ in vortex formation. The calculation has been performed for $R=0\ldots 20c/\omegd$ for $\omegd r_{\rm b}/c=0.43$. Thin vertical lines show the scale of the interaction potential, $|x_1-x_2|=\pm r_{\rm b}$.
%\textbf
{(e-g) Three-photon phase profiles:} The phase of the three-photon wavefunction under the single-band model (e) retains a 6-fold rotational symmetry and fails to capture the warping. The multiband model (f) reduces the rotational symmetry to 3-fold, with the delay $\Delta R$ manifesting here as a delay of the vortex formation along the pair-ahead (red) lines. In (g), the isosurfaces of $|\nabla\mathop\mathrm{arg}(\psi)|$ reveal a vortex ring with trigonal warping, a hallmark of the three-photon interaction captured by the multiband model. 
Calculations in (e,f,g) follow Eqs.~\eqref{eq:K3c}--\eqref{eq:K3d}, with $\omegd r_{\rm b}/c=1$. Panels (e,f) correspond to {$R=7.6c/\omegd$}.
}\label{fig:Dirac}
\end{figure*}

In a homogeneous medium, Eqs.~\eqref{eq:two} and \eqref{eq:three} remain independent of the normalized center-of-mass coordinate $R=\sum_{i=1}^nx_i/\sqrt{n}$. 
Thus, it is useful to introduce the normalized relative coordinates: $r=(x_1-x_2)/\sqrt{2}$ for $n=2$ and the spatial Jacobi coordinates $\upeta=x_{12}/\sqrt{2}$ and 
$\upzeta=(x_{13}+x_{23})/\sqrt{6}$ for $n=3$, with $x_{ij}=x_i-x_j$.
Before discussing interaction effects between polaritons, we examine the spatial dispersion outside the blockade region. 
To this end, we assume translational invariance and replace the spatial derivatives $-\rmi\partial_R$,
{$-\rmi\partial_r$}, $-\rmi\partial_{\upeta,\upzeta}$ by the corresponding  momenta $K$, $k$, $k_{\upeta,\upzeta}$. This allows Eqs.~\eqref{eq:two} and \eqref{eq:three} to be solved for the normalized center-of-mass momentum $K$ as a function of $k$ and $\vec{k}=(k_\upeta,k_\upzeta)$.  

For two photons, Eqs.~\eqref{eq:two} yield
 \begin{equation}
    -\frac{\rmi  }{\sqrt{2}}\frac{\partial}{\partial R}\psi=\frac{1}{\sqrt{2}}
    \begin{pmatrix}
        -k&0\\0&k
    \end{pmatrix} \psi
    {-}\frac{{\omegd}}{c}\psi{+}\frac{\omegd}{2c}\begin{pmatrix}
     1&1\\1&1
     \end{pmatrix}
       \psi\:,\label{eq:K2}
\end{equation}
where $\omegd\equiv V_2(\infty)=\rho g^2/\Delta$. Replacing $-\rmi\partial _R$ by $K$
and diagonalizing the matrix, we obtain
\begin{equation}\label{eq:K2b}
K_\pm(k)=
-\frac{\sqrt{2}{\omegd}}{2c}
\pm \sqrt{\frac1{2}\left(\frac{\omegd}{c}\right)^2+k^2}\:.
\end{equation}
This characteristic Dirac dispersion is plotted in Fig.~\ref{fig:dispersion}a.
At $k=0$, there is a gap, and the eigenstates are the even and odd superpositions, $ES_\pm=(ES\pm SE)/\sqrt{2}$. Near $k=0$, the dispersion is parabolic (dotted curve in Fig.~\ref{fig:dispersion}a), corresponding to the Schr\"odinger  approximation~\cite{Peyronel2012,firstenberg2013attractive},
\begin{equation}\label{eq:Shr}
    \rmi \frac{\partial}{\partial R} ES_+=
    \biggl(-\frac{1}{2m}\frac{\partial^2}{\partial r^2}{+\frac{\sqrt{2}{\omegd}}{c}-\frac{\sqrt{2}V_2}{c}}\biggr)ES_+~,
\end{equation}
where  $m=-\omegd/(\sqrt{2}c) $  is the effective mass. For large $k$, the dispersion becomes linear and deviates from the Schr\"odinger approximation.

The difference between the parabolic approximation and the full model is more pronounced for three polaritons, as shown in Fig.~\ref{fig:dispersion}b. To derive the multiband structure, we rewrite Eq.~\eqref{eq:three} using the relative momenta $k_{\upeta,\upzeta}$ and center-of-mass coordinate $R$,
\begin{equation}
\begin{split}
    -\frac{\rmi  }{\sqrt{3}}\frac{\partial}{\partial R}\psi=
    \frac{\sqrt{6}}{3}\begin{pmatrix}
        -\frac{1}{2}k_{\upzeta}-\frac{\sqrt{3}}{2}k_{\upeta}&0&0\\0&-\frac{1}{2}k_{\upzeta}+\frac{\sqrt{3}}{2}k_{\upeta}&0\\0&0&k_{\upzeta}
    \end{pmatrix} \psi
    \\{-}\frac{{\omegd}}{c}\psi{+}\frac{\omegd}{3c}\begin{pmatrix}
     1&1&1\\1&1&1\\1&1&1
     \end{pmatrix}
       \psi\:,
       \end{split}\label{eq:K3}
\end{equation}
and transform to the basis  $ESS_{+}=(ESS+ SES+SSE)/\sqrt{3}$, $ESS_{-}=(ESS- SES)/\sqrt{2}$, $ESS_{-}'=(ESS+ SES-2SSE)/\sqrt{6}$, 
\begin{equation}
\begin{split}
      -\frac{\rmi  }{\sqrt{3}}\frac{\partial}{\partial R}\psi=
    -\frac{1}{\sqrt{3}}\begin{pmatrix}
        0&k_{\upeta}&k_{\upzeta}\\
        k_{\upeta}&k_{\upzeta}/\sqrt{2}&k_{\upeta}/\sqrt{2}\\
        k_{\upzeta}&k_{\upeta}/\sqrt{2}&-k_{\upzeta}/\sqrt{2}\\
    \end{pmatrix} \psi
    \\{-}\frac{{\omegd}}{c}\psi{+}\frac{\omegd}{c}\begin{pmatrix}
     1&0&0\\0&0&0\\0&0&0
     \end{pmatrix}
       \psi\:.
       \end{split}\label{eq:K3b}
\end{equation} 
At $\vec{k}=0$, solving Eq.~\eqref{eq:K3b} for $K$ yields one symmetric state, $ESS_{+}$ (blue surface in Fig.~\ref{fig:dispersion}b), and two degenerate antisymmetric states, $ESS_{-}$ and $ESS_{-}'$ (red surfaces in Fig.~\ref{fig:dispersion}b).  Near $\vec{k}=0$, the symmetric state $ESS_+$ acquires mass,
\begin{equation}\label{eq:m3}
K(\vec{k})=
\frac{1}{\sqrt{3}}\frac{ck^2}{\omegd},
\end{equation}
while the two antisymmetric states split, forming characteristic Dirac cones described by
\begin{equation}\label{eq:m3_degen}
K=-\sqrt{3}\frac{{\omegd}}{c}\pm \frac{|k|}{\sqrt{2}}, \quad \mathrm{where}~k\equiv\sqrt{k_\upeta^2+k_\upzeta^2}\:.
\end{equation}

Far from $\vec{k}=0$, all three states mix, yielding warped dispersion with $C_{3v}$ symmetry. This effect is similar to the trigonal warping of Dirac cones in graphene~\cite{Bradlyn2016}, where warping arises from the crystalline symmetry. Here, it emerges naturally from the three-polariton coupling and permutation symmetry. Importantly, this warping cannot be captured by the parabolic  equation Eq.~\eqref{eq:m3}, which approximates the warped cone as a rotationally symmetric paraboloid (black surface in Fig.~\ref{fig:dispersion}b). 

We note that Eqs.~\eqref{eq:two} and \eqref{eq:three} that rely on $|\mathrm{p}\rangle$-state elimination retain the minimal number of bands sufficient to capture the symmetry. We provide a comparison to the full model -- without state eliminations -- in Appendix \ref{app_full_model} and Figs.~\ref{fig:two-photon-dispersion} and \ref{fig:three-photon-dispersion}. Figure~\ref{fig:three-photon-dispersion} demonstrates for $n=3$ photons that the full model, comprising $3^3=27$ components, features 19 bands and two degenerate Dirac-like points, resulting from the splitting of the Dirac point in Fig.~\ref{fig:dispersion}(b). This splitting 
can be seen as a fine structure of the model Eq.~\eqref{eq:three}, that is sensitive to the contribution of $|\mathrm{p}\rangle$ states and appears in the next order in $\Omega/\Delta$. Our analysis shows that the minimal model Eq.~\eqref{eq:three} very well describes the dispersion of the experimentally relevant $ESS_+$ band, shown by the blue warped cone in Fig.~\ref{fig:Dirac}(b).

\subsection{Real-space consequences}
\noindent We now demonstrate how the difference between the single-band Schr\"odinger and multiband Dirac-like dispersions manifests in the real-space wavefunctions. For simplicity, we consider a uniform medium of length $\len$ (one typically chooses $\len\approx \sqrt{2\pi}\sigma$ for a Gaussian cloud with width $2\sigma$  \cite{DroriScience2023}).
For two photons, it is instructive to divide the Hilbert space into quadrants corresponding to zero, one, or two photons inside the atomic medium, as shown in Fig.~\ref{fig:Dirac}a. Photon-photon interaction, mediated by the atoms, occurs only when both photons are inside (\textit{i.e.}, in the gray-shaded quadrant $\zero<x_{1,2}<\len$). The green lines $x_1=\zero$ and $x_2=\zero$, marking the entry points of the first and second photons, define the boundary conditions for the onset of interaction. These lines serve as effective light-cone boundaries for the development of correlations.
Nevertheless, to simplify semi-analytic modeling, the boundary conditions are typically adjusted so that the interaction begins earlier, at $R=\zero$ for all $r$ (dashed line in Fig.~\ref{fig:Dirac}a)~\cite{firstenberg2013attractive,DroriScience2023}. Initially uncorrelated photons are represented by the wavefunction $\psi(R=\zero,r)=1$, which accumulates phase as it propagates.

The steady-state wavefunction in the single-band Schr\"odinger approximation is shown in Fig.~\ref{fig:Dirac}c.
 Phase accumulates more rapidly when photons are close together, generating vortex-antivortex pairs. The generation mechanism is robust, as it results from interference between the two-photon bound state and the unbound part of the wavefunction \cite{DroriScience2023}. 
While the single-band model predicts these vortices, it also produces unphysical correlations below the light-cone boundaries $R=|r|$ (i.e. at $x_{1,2}<0$, where only one photon is inside). These arise because the Schr\"odinger equation does not impose a speed limit on information propagation. 

By contrast, our dual-band Dirac-like model, Eq.~\eqref{eq:two}, naturally enforces the light-cone limits, as seen in Fig.~\ref{fig:Dirac}d. Furthermore, since two-photon interaction starts only at $R=|r|$ in the dual-band model, the vortices shift forward by $\Delta R$. This shift can thus be understood as a consequence of the light-cones limits imposed on the phase pattern in Fig.~\ref{fig:Dirac}d. 

\begin{figure*}
\centering\includegraphics[width=0.84 \textwidth]{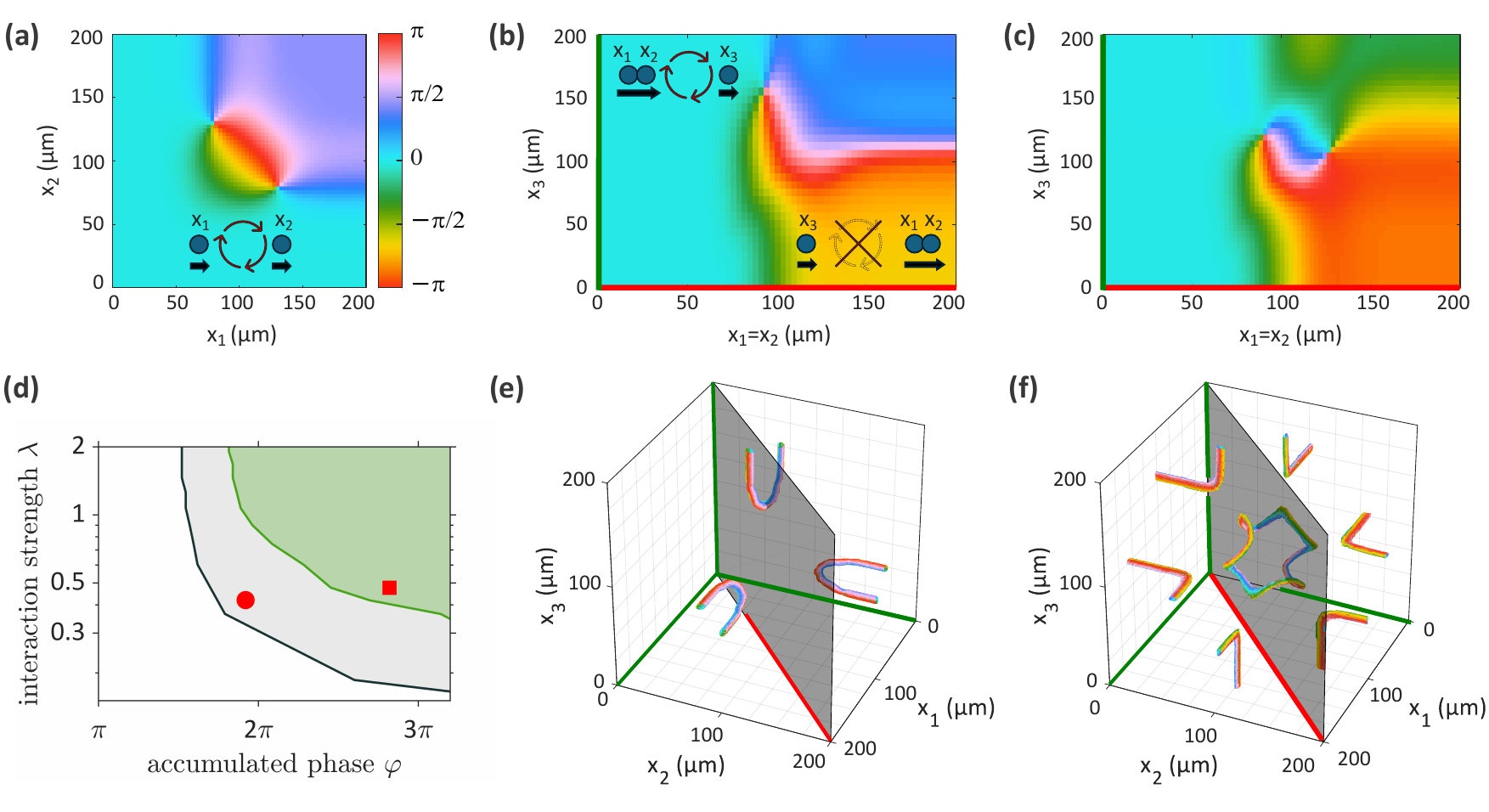}
\caption{%\textbf
{Numerical simulations of three-photon vortices with three-fold rotational symmetry $C_{3v}$.} A Gaussian-shaped atomic cloud is centered at $x=100~\mu$m ($1\sigma=30~\mu$m, $\OD=78-115$). %\textbf
{(a)} Phase of the stationary two-photon wavefunction $\psi(x_1,x_2)$, showing a symmetric vortex-antivortex pair. %\textbf
{(b,c,e,f)} Phase of the stationary three-photon wavefunction $\psi(x_1,x_2,x_3)$, displaying different vortex tube and ring configurations for (b,e) intermediate and (c,f) strong interactions. (e,f) show isosurfaces of $|\nabla\mathop\mathrm{arg}(\psi)|$, while (b,c) present cross-sections of $\mathop\mathrm{arg}(\psi)$ along the shaded planes in (e,f). For an intermediate interaction strength (b,e), three-photon vortex tubes form only in the single-ahead configurations (e.g., $x_1\approx x_2< x_3$), consistent with faster pair propagation. For longer and stronger interactions (c,f), vortex tubes also form in the pair-ahead configurations (e.g., $x_3<x_1\approx x_2$), and the six tubes merge into a warped vortex ring with a reduced $C_{3v}$ symmetry. 
%\textbf
{(d) Phase diagram of three-photon vortex formation.} The shaded regions mark the parameter regimes for (gray) single-ahead and (green) pair-ahead vortices as a function of interaction strength ($\lambda$) and duration ($\varphi$). 
The red circle and square correspond to the conditions of (b,e) and (c,f), respectively. Calculations here follow Eqs.~(\ref{eq:3_photon_equations}), with
the following parameters: $\sqrt{2\pi}\sigma=75$ $\mu \rm m$, $r_\mathrm{b}=15.3$ $\mu \rm m$, $\Omega=9.5$ MHz, $\Gamma= 3.03$ MHz, $\gamma=0.07$ MHz, $\Delta=28.5$ MHz, $\delta=1.03$ MHz (note that decay rates and Rabi frequencies are given in the half-width convention throughout this paper). The optical depths are:
OD= 90 ($\varphi=2.21\pi$, $\lambda=2.03$) for Fig.~\ref{fig:Chiral}a, OD= 78 ($\varphi=1.92\pi$, $\lambda=0.42$) for Figs.~\ref{fig:Chiral}b,e, and OD= 115 ($\varphi=2.82\pi$, $\lambda=0.47$) for Figs.~\ref{fig:Chiral}c,f. 
}
\label{fig:Chiral}
\end{figure*}

For three photons, the impact of the multiband dispersion is more pronounced and includes symmetry reduction. We first analyze this effect analytically. The space $(x_1,x_2,x_3)$ is divided into sections based on the number of photons inside the medium (Fig.~\ref{fig:Dirac}b). The faces of the three-photon section $\zero<x_{1,2,3}<\len$ (gray shaded) act as effective light-cone boundaries, warping the wavefunction. To understand the warping symmetry, we identify two types of regions in that section: Three `single ahead' regions, where a single photon propagates ahead of a tight photon pair (\textit{e.g.}, $x_1\approx x_2< x_3$, or $\upzeta<0$ for $\upeta\approx0$), and three `pair ahead' regions, where a single photon lags behind a pair (\textit{e.g.}, $x_3<x_1\approx x_2$, or $\upzeta>0$ for $\upeta\approx0$). The single-ahead regions of the wavefunction evolve predominantly from the section edges (green lines in Fig.~\ref{fig:Dirac}b,e-g), while the pair-ahead regions evolve from the face centers (red lines). 

As evident from the geometry, single-ahead configurations begin to experience three-photon interaction at an `earlier' $R=(x_1+x_2+x_3)/\sqrt{3}$ compared to pair-ahead configurations. They develop stronger correlations during propagation, further reducing the wavefunction symmetry. 
This effect is amplified by the group velocity dependence on photon-photon interaction. Since a closely spaced photon pair moves faster than a single photon \cite{firstenberg2013attractive,Bienias2014,Chang2022}, the single-ahead configurations gradually compact and interact more, while in pair-ahead configurations, the single photon lags behind.

This behavior is captured by the multiband model Eq.~\eqref{eq:three} in numerical wavefunction evolution (Fig.~\ref{fig:Dirac}f). It yields a reduced $C_{3v}$ symmetry, with correlations building up faster for single-ahead compared to pair-ahead configurations, as evident in Fig.~\ref{fig:Dirac}f by the greater accumulated conditional phase along the green lines relative to the red lines.
In contrast, the single-band model retains the $C_{6v}$ symmetry (Fig.~\ref{fig:Dirac}e). The three-photon interaction ultimately produces a phase vortex ring \cite{DroriScience2023}, which the multiband model predicts to exhibit a trigonal warping (Fig.~\ref{fig:Dirac}i). 
We further explore this warping in the next section. 

\subsection{Trigonal warping of three-photon vortices}
We have developed a full numerical model to simulate the exact dynamics in a Gaussian-shaped atomic cloud, extending previous two-photon models \cite{Gorshkov2011Sep,Peyronel2012}. 
 We explore the formation of three-photon vortices by numerically solving all the 27 propagation equations [Eqs.~(\ref{eq:3_photon_equations})], as shown in Fig.~\ref{fig:Chiral}. 
 This provides the complete three-photon wavefunction $\psi(x_1,x_2,x_3)$ with its $3^3=27$ components inside the cloud.
  For reference, Fig.~\ref{fig:Chiral}a displays the vortex-antivortex pair in the two-photon wavefunction. A vortex is always accompanied by an antivortex due to the bosonic symmetry, $\psi(x_1,x_2)=\psi(x_2,x_1)$.
For the three-photon wavefunction $\psi(x_1,x_2,x_3)$, although the $C_{3v}$ bosonic symmetry still holds, the single-ahead ($x_1\approx x_2< x_3$) and pair-ahead ($x_3<x_1\approx x_2$) configurations exhibit markedly different behavior. 

At intermediate interaction strengths (Fig.~\ref{fig:Chiral}b), vortices form only in the single-ahead configuration.  
Increasing the interaction enables vortices in both configurations (Fig.~\ref{fig:Chiral}c), though the wavefunction remains asymmetric. The phase diagram in Fig.~\ref{fig:Chiral}d maps the parameter regimes for forming single-ahead vortices (gray shading) and all vortices (green shading). We quantify the interaction strength and duration using the dimensionless parameters $\lambda\propto{\rm OD}_\mathrm{b}^2$ and $\varphi\propto{\rm OD}$, respectively \cite{DroriScience2023} (see Appendix \ref{app_phase_diagram_params} for exact expressions). Here, OD is the optical depth of the medium, and ${\rm OD}_\mathrm{b}={\rm OD}\cdot r_{\rm b}/\len$ is the optical depth of the blockade range. Similar to the threshold for two-photon vortices \cite{DroriScience2023}, we identify thresholds for both types of three-photon vortices (black and green lines).

Figures \ref{fig:Chiral}(e,f) illustrate the corresponding three-dimensional vortex structures. The first regime (Fig.~\ref{fig:Chiral}e) is characterized by three disconnected vortex tubes, each originating from a single-ahead vortex. Their trigonal symmetry, their emergence only when the three photons are inside the medium, and their curved trajectories---all signal the presence of genuine three-body interactions.

In the second regime (Fig.~\ref{fig:Chiral}f), three pairs of vortex tubes emerge already from pair-ahead vortices before the third photon enters, reflecting two-body interactions with a six-fold symmetry. As the third photon joins, these tubes merge into a central three-body vortex ring. The warped ring clearly reflect the trigonal symmetry of the underlying dispersion and the light-cone boundaries.

\section*{Multiphoton model}
\noindent  The simplified multiband model, Eqs.~\eqref{eq:two} and \eqref{eq:three}, extends naturally to $n>3$ photons. For instance, for $n=4$, the Hamiltonian generalizes straightforwardly to:
\begin{multline}\label{eq:H4}
    \mathcal{H}(x_1\ldots x_4)=  H(x_1)\otimes I\otimes I\otimes I \\+  I\otimes H(x_2)\otimes I\otimes I+\ldots +I\otimes I\otimes I
    \otimes H(x_4)\:
\end{multline}
and $\psi=[ESSS,SESS,SSES,SSSE]$ after eliminating the $SSSS$ state. The general result is a system of $n$ equations
\begin{equation}\label{eq:n}
       \rmi\frac{\partial \psi_i}{\partial t}={-}\rmi c  \frac{\partial \psi_i}{\partial x_i}
+{{\omegd}}\psi_i+\frac{1}{n}V_n(x_1,\ldots x_n)       \sum\limits_{i=1}^n\psi_{i}\:,
    \end{equation}
 where $\psi_i$ is the wavefunction amplitude with the $i$-th polariton in state $E$ and the others in state $S$, and $V_n$ is the rescaled vdW interaction potential {Eq.~\eqref{eq:Vn}}.

In the blockade range, where any two polaritons are close, $V_n=0$. 
Without the interaction term $V_n\sum_{i=1}^n\psi_{i}$,  Eq.~\eqref{eq:n} describes $n$ free photons, each propagating independently with linear (massless) dispersion, 
$\psi_i\propto\e^{-\rmi\omega_{k_i}t+\rmi k_ix_i}$\:,
where $\omega_k=ck$, $k_i$ is the wave vector of $i$-th photon, and we assume the translation invariance.            
Since the photons are hybridized with the Rydberg state, the coupling term persists even at large separations, where $V_n$ approaches the constant $V_{n,\infty}=\omegd$. 
        
Following the approach for $n=2,3$, we solve Eq.~\eqref{eq:n} in a translationally invariant system for the normalized center-of-mass momentum~ $K=\sum_{i=1}^nk_i/\sqrt{n}$. 
We identify one massive mode with all $\psi_i=1/\sqrt{n}$ and $n-1$ degenerate modes with $\sum_{i=1}^n\psi_i=0$. 
The massive mode exhibits a parabolic dispersion consistent with the Schr\"odinger approximation~\cite{Peyronel2012,firstenberg2013attractive}. 
However, for large $k$, the dispersion becomes linear, deviating from the Schr\"odinger model. The degenerate modes split linearly with $k_i$ for $k_i\ne 0$.

We now illustrate this in the stationary case for $n=4$ photons and assume that they are sufficiently far apart so that $V_n\equiv \omegd.$ Using the center-of-mass reference frame and the generalized Jacobi coordinates  \cite{LEROY1987}, we define $R=\sum_{i=1}^n x_i/2, \eta_{1}=(x_{1}-x_{2})/\sqrt{2}, \eta_{2}=(x_{3}-x_{4})/\sqrt{2}$, and $\eta_{3}=(x_{1}+x_{2}-x_{3}-x_{4})/2$.
The four-photon dispersion equation then becomes
\begin{align}       \label{eq:K4}
    -\frac{\rmi  }{2}&\frac{\partial}{\partial R}\psi\\&=
\left(\begin{smallmatrix}
\frac{1}{\sqrt{2}}\kappa_{1}+\frac{1}{2}\kappa_{3}&0&0&0\\        
0&-\frac{1}{\sqrt{2}}\kappa_{1}+\frac{1}{2}\kappa_{3}&0&0\\
        0&0&\frac{1}{\sqrt{2}}\kappa_{2}-\frac{1}{2}\kappa_{3}&0\\
                0&0&0&-\frac{1}{\sqrt{2}}\kappa_{2}-\frac{1}{2}\kappa_{3}
    \end{smallmatrix}\right) \psi\nonumber
    \\&{-}\frac{{\omegd}}{c}\psi{+}\frac{\omegd}{4c}\begin{pmatrix}
     1&1&1&1\\1&1&1&1\\1&1&1&1\\1&1&1&1
     \end{pmatrix}
       \psi\:,\nonumber
\end{align}
where the relative motion wavevectors are $\kappa_{i}=-\rmi\partial_{\eta_{i}}$ and $-\rmi \partial_R\equiv K$. The massive eigenstate of Eq.~\eqref{eq:K4} corresponds to
\begin{equation}\label{eq:K4b}
K=\frac{c\kappa^{2}}{2\omegd}\:,
\end{equation}
{where $\kappa^2\equiv \sum_{i=1}^n\kappa_i^2$\:.}

\begin{figure}[t]
\centering\includegraphics[width=0.3\textwidth]{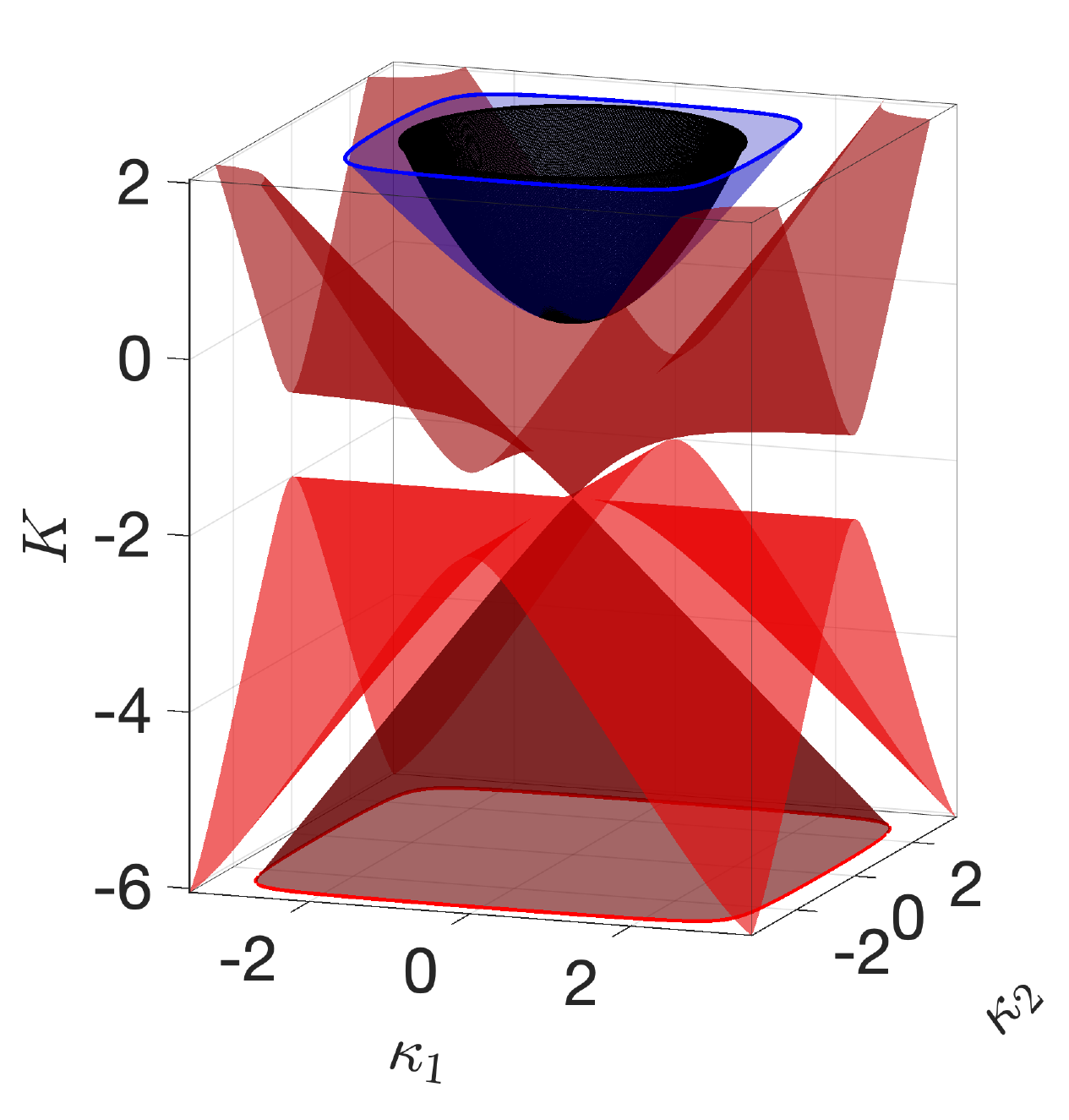}
\caption{
{Analytical multiband model for $n=4$ photons.} The surfaces show
momentum-space dispersion calculated following Eq.~\eqref{eq:K4} for $\kappa_3=0$.
Shaded paraboloid shows the Schr\"odinger approximation result, Eq.~\eqref{eq:K4b}. 
All units are $\omegd$.
}\label{fig:SN4}
\end{figure}

The corresponding four-photon dispersion is shown in Fig.~\ref{fig:SN4}. The dispersion exhibits similarities to Fig.~\ref{fig:dispersion}b  for $n=3$ photons. However, in this case, we observe a warped cone (blue) with a 4-fold rotational symmetry rather than a 3-fold one. Additionally, a more complex three-fold degeneracy (red bands) emerges, which splits for $\kappa<0$.  Such a dispersion law has similarities with a chiral dispersion analogous to chiral fermions in three-dimensional crystals~\cite{Bradlyn2016}.
This analysis lays the groundwork for future studies of $n>3$ polaritons.

\section*{Discussion and outlook}
To summarize, we have analyzed how the spatial symmetry of the wavefunction $\psi(x_1, \ldots, x_n)$ of $n$ interacting polaritons emerges from its multicomponent structure. We provide analytical models and solutions for multi-polariton dispersion and spatial propagation, as well as a full numerical model for three photons that describes the evolution in a nonuniform medium and the resulting temporal correlations.  

Our model predicts $C_{3v}$ symmetry of warped vortex rings present in the phase of $\psi(x_1, x_2, x_3)$.  This symmetry arises from the interplay between the multiband nature of the dispersion and the medium's finite spatial extent. Both ingredients are essential for its emergence in the steady state.
Thus,  strongly interacting polaritons should manifest effects not captured by the traditionally accepted models of interacting particles with parabolic dispersion and contact interactions ~\cite{GullansPRL2016,Liang2018}. 

The existence of high-order dispersion degeneracies for chiral fermions in three-dimensional solids, extending beyond traditional Dirac or Weyl models, is well established~\cite{Orlita_2014,Bradlyn2016,Rao_2019}. Here, we show that analogous degeneracies, potentially in even higher dimensions, are intrinsic to the multicomponent interacting polaritons, leading to observable phenomena. We anticipate further fundamental insights in other excitation geometries, such as counter-propagating polaritons. It is also potentially interesting to examine such  exotic
 few-body effects in other quantum platforms where several polaritons propagate in one dimension, for example, for waveguide-coupled atoms~\cite{sheremet2023} or Rydberg superatoms~\cite{Hofferberth2018,Kumlin_2023}.

\section*{Acknowledgments}
We thank Leonid Golub, Ashley Harkavi, and Aditya Prakash for useful discussions. 
We acknowledge financial support from the Israel Science Foundation (grant No.~3491/21, 1982/22), the US-Israel Binational Science Foundation and US National Science Foundation, the Minerva Foundation with funding from the Federal German Ministry for Education and Research, the Leona M.~and Harry B.~Helmsley Charitable Trust, the Shimon and Golde Picker - Weizmann Annual Grant, and the Laboratory in Memory of Leon and Blacky Broder.

\vspace{0.5cm}

\section*{APPENDICES}
\vspace{0.5cm}

\appendix

\section{Numerical model for three interacting polaritons}\label{app_numerical_model}

\noindent For the numerical model, we consider a weak coherent state of light, $\psi(t)$, impinging on the medium of three-level atoms. The total wavefunction is given by
\begin{widetext}
\begin{equation}\label{eq:psi}
\begin{array}{lll}
    \ket{\psi(t)}&=& \epsilon \ket{0}+ \int\limits_{-\infty}^{+\infty}\sum\limits_{i=1}^{3}dx_1 A_i(x_1,t)\hat{a_i}^\dagger(x_1)\ket{0}+
   \dfrac{1}{2} \iint\limits_{-\infty}^{+\infty} \sum\limits_{i,j=1}^{3}dx_1dx_2A_{i}A_{j}(x_1,x_2,t)\hat{a_i}^\dagger(x_1)\hat{a_j}^\dagger(x_2)\ket{0}\\
   &&+
   \dfrac{1}{3!} \iiint\limits_{-\infty}^{+\infty} \sum\limits_{i,j,k=1}^{3}dx_1dx_2dx_3A_{i}A_{j}A_{k}(x_1,x_2,x_3,t)\hat{a_i}^\dagger(x_1)\hat{a_j}^\dagger(x_2)\hat{a_k}^\dagger(x_3)\ket{0},
 \end{array}
\end{equation}
\end{widetext}
where $A=[E,P,S]$, and the excitation operators are $\hat{a}=[\hat{\varepsilon},\hat{p},\hat{s}]$. We assume $|\epsilon|\ll 1$ and neglect four or more excitations.  
As described in the main text, the ground $\ket{g}$, excited $\ket{p}$, and Rydberg $\ket{s}$ states are coupled via the level-type scheme
\[\Ket{g} \overset{g, \sqrt{\rho}}{\underset{\Delta,\Gamma}\rightleftarrows} \ket{p}\overset{\Omega}{\underset{\delta,\gamma}\rightleftarrows}\ket{s},\]
forming Rydberg polaritons. 
The Hamiltonian governing the single-polariton propagation $H(x)\psi=0$, where $\psi(x)=[E(x),P(x),S(x)]^T$, is given by Eq.~\eqref{eq:H} with parameters defined therein.

The two- and three-polariton wavefunctions $\psi(x_1,x_2)$ and $\psi(x_1,x_2,x_3)$ have $3^2$ and $3^3$ components, respectively. The effective Hamiltonian for a free three-polariton propagation is thus given by a 
 27$\times$27 matrix constructed from $H(x)$ as
\begin{equation}\label{eq:H3}
    \mathcal{H}(x_1,x_2,x_3)= I \otimes H(x_2)\otimes I + I \otimes I\otimes H(x_3)+ H(x_1)\otimes I\otimes I\:,
\end{equation}
where  $I$ is the $3\times 3$ identity matrix. 
Additionally, the Rydberg interaction between the polaritons is described by the Hamiltonian 
 \begin{equation}\label{eq:V}
    \mathcal{V}= \dfrac{1}{2}\iint dx dx' V_\mathrm{ss}(x,x')\hat{S}^\dagger(x')\hat{S}^\dagger(x)\hat{S}(x)\hat{S}(x'),
\end{equation}
where $\hat{S}(x)$ is the excitation operator to the Rydberg state, and 
 $V_\mathrm{ss}(x,x')= C_6/|x-x'|^6$ is the Rydberg interaction potential.
In our calculations, we project this Hamiltonian onto the three-polariton space.

We consider an atomic cloud with a Gaussian density profile $\rho=\rho(x)$ centered at $\xo/2=5\sigma$, such that the range $x\in [0,\xo]$ fully contains the cloud. The equations of motion for the three-photon amplitudes then read:
\begin{widetext}

\begin{align}\label{eq:3_photon_equations}
 \partial_t EEE&= -c(\partial_{x_1}+\partial_{x_2}+\partial_{x_3})EEE + i \tilde{g}(x_1)PEE+ i \tilde{g}(x_2)EPE+ i \tilde{g}(x_3)EEP \nonumber \:,\\  
     \partial_t EEP &= -c(\partial_{x_1}+\partial_{x_2})EEP -(\Gamma- i\Delta)EEP + i \tilde{g}(x_1)PEP+ i \tilde{g}(x_2)EPP+ i \tilde{g}(x_3)EEE+ i \Omega EES \nonumber \:,\\ 
    \partial_t EES&= -c(\partial_{x_1}+\partial_{x_2})EES -(\gamma- i\delta)EES + i \tilde{g}(x_1)PES+ i \tilde{g}(x_2)EPS+ i \Omega EEP \nonumber \:,\\ 
 \partial_t EPE&= -c(\partial_{x_1}+\partial_{x_3})EPE -(\Gamma- i\Delta)EPE + i \tilde{g}(x_1)PPE+ i \tilde{g}(x_2)EEE+ i \tilde{g}(x_3)EPP+ i \Omega ESE \nonumber \:,\\
 \partial_t EPP&= -c\partial_{x_1}EPP -2(\Gamma- i\Delta)EPP + i \tilde{g}(x_1)PPP+ i \tilde{g}(x_2)EEP+ i \tilde{g}(x_3)EPE+ i \Omega EPS+ i  \Omega ESP \nonumber \:,\\
 \partial_t EPS&= -c\partial_{x_1}EPS -(\Gamma- i\Delta)EPS -(\gamma- i\delta)EPS+ i \tilde{g}(x_1)PPS+ i \tilde{g}(x_2)EES+ i \Omega EPP+ i \Omega ESS \nonumber \:,\\
 \partial_t ESE&= -c(\partial_{x_1}+\partial_{x_3})ESE -(\gamma- i\delta)ESE+ i \tilde{g}(x_1)PSE+ i \tilde{g}(x_3)ESP+ i\Omega EPE \nonumber \:,\\
\partial_t ESP&= -c\partial_{x_1}ESP -(\Gamma- i\Delta)ESP -(\gamma- i\delta)ESP+ i \tilde{g}(x_1)PSP+ i \tilde{g}(x_3)ESE+ i  \Omega EPP+ i \Omega ESS \nonumber \:,\\
\partial_t ESS&= -c\partial_{x_1}ESS -2(\gamma- i\delta)ESS+ i \tilde{g}(x_1)PSS+ i \Omega EPS+ i \Omega ESP -i V_{ss}(x_2,x_3)ESS \nonumber\:, \\
   \partial_t PEE&= -c(\partial_{x_2}+\partial_{x_3})PEE -(\Gamma- i\Delta)PEE+ i \tilde{g}(x_1)EEE+i \tilde{g}(x_2)PPE+i \tilde{g}(x_3)PEP+i  \Omega SEE \nonumber \:,\\  
     \partial_t PEP&= -c\partial_{x_2}PEP -2(\Gamma- i\Delta)PEP+ i \tilde{g}(x_1)EEP+i\tilde{g}(x_2)PPP+i \tilde{g}(x_3)PEE+i  \Omega PES+i \Omega SEP \nonumber \:,\\ 
   \partial_t PES&= -c\partial_{x_2}PES -(\Gamma- i\Delta)PES-(\gamma- i\delta)PES+ i \tilde{g}(x_1)EES+i \tilde{g}(x_2)PPS+i  \Omega PEP+i \Omega SES \nonumber \:,\\
   \partial_t PPE&= -c\partial_{x_3}PPE -2(\Gamma- i\Delta)PPE+ i \tilde{g}(x_1)EPE+i \tilde{g}(x_2)PEE+i \tilde{g}(x_3)PPP+i \Omega PSE+i \Omega SPE \nonumber \:,\\
   \partial_t PPP&=  -3(\Gamma- i\Delta)PPP+ i \tilde{g}(x_1)EPP+i \tilde{g}(x_2)PEP+i \tilde{g}(x_3)PPE+i \Omega PPS+i \Omega PSP+i  \Omega SPP \nonumber \:,\\
   \partial_t PPS&=  -2(\Gamma- i\Delta)PPS-(\gamma- i\delta)PPS+ i \tilde{g}(x_1)EPS+i \tilde{g}(x_2)PES+i  \Omega PPP+i  \Omega PSS+i \Omega SPS \nonumber \:,\\
   \partial_t PSE&=  -c\partial_{x_3}PSE -(\Gamma- i\Delta)PSE-(\gamma- i\delta)PSE+ i \tilde{g}(x_1)ESE+i \tilde{g}(x_3)PSP+i \Omega PPE+i  \Omega SSE \nonumber \:,\\
   \partial_t PSP&= -2(\Gamma- i\Delta)PSP-(\gamma- i\delta)PSP+ i \tilde{g}(x_1)ESP+i \tilde{g}(x_3)PSE+i  \Omega PPP+i  \Omega PSS+i  \Omega SSP \nonumber \:,\\
   \partial_t PSS&= -(\Gamma- i\Delta)PSS-2(\gamma- i\delta)PSS+ i \tilde{g}(x_1)ESS+i  \Omega PPS+i  \Omega PSP+i \Omega SSS-iV_\mathrm{ss}(x_2,x_3)PSS \nonumber \:,\\
   \partial_t SEE&= -c(\partial_{x_2}+\partial_{x_3})SEE-(\gamma- i\delta)SEE+ i \tilde{g}(x_2)SPE+ i \tilde{g}(x_3)SEP +i  \Omega PEE \nonumber \:,\\
   \partial_t SEP&= -c\partial_{x_2}SEP -(\Gamma- i\Delta)SEP-(\gamma- i\delta)SEP+ i \tilde{g}(x_2)SPP+ i \tilde{g}(x_3)SEE+i  \Omega PEP+i \Omega SES \nonumber \:,\\
   \partial_t SES&= -c\partial_{x_2}SES -2(\gamma- i\delta)SES+ i \tilde{g}(x_2)SPS +i \Omega PES+i \Omega SEP -iV_\mathrm{ss}(x_1,x_3)SES \nonumber \:,\\
   \partial_t SPE&= -c\partial_{x_3}SPE -(\Gamma- i\Delta)SPE -(\gamma- i\delta)SPE+ i \tilde{g}(x_2)SEE + i \tilde{g}(x_3)SPP +i  \Omega PPE+i  \Omega SSE \nonumber \:,\\
   \partial_t SPP&=  -2(\Gamma- i\Delta)SPP -(\gamma- i\delta)SPP+ i \tilde{g}(x_2)SEP + i \tilde{g}(x_3)SPE+i \Omega PPP+i \Omega SPS +i\Omega SSP \nonumber \:,\\
   \partial_t SPS&=  -(\Gamma- i\Delta)SPS -2(\gamma- i\delta)SPS+ i \tilde{g}(x_2)SES +i  \Omega PPS+i\Omega SPP +i\Omega SSS -iV_\mathrm{ss}(x_1,x_3)SPS \nonumber \:,\\
   \partial_t SSE&= -c\partial_{x_3}SSE-2(\gamma- i\delta)SSE+ i \tilde{g}(x_3)SSP+i  \Omega PSE+i  \Omega SPE-iV_\mathrm{ss}(x_1,x_2)SSE \nonumber \:,\\
   \partial_t SSP&= -(\Gamma- i\Delta)SSP-2(\gamma- i\delta)SSP+ i \tilde{g}(x_3)SSE +i \Omega PSP+i \Omega SPP +i \Omega SSS -iV_\mathrm{ss}(x_1,x_2)SSP \nonumber \:,\\
   \partial_t SSS&= -3(\gamma- i\delta)SSS+ i \Omega PSS+i \Omega SPS +i \Omega SSP -iV_\mathrm{sss}(x_1,x_2,x_3)SSS \:.
\end{align}
\end{widetext}
Here, the three-body interaction term is defined as $V_\mathrm{sss}
(x_1,x_2,x_3)=\sum_{i\neq j} V_\mathrm{ss}(x_i,x_j)$, and the spatially dependent coupling is given by $\tilde{g}(x)=g\sqrt{\rho(x)}$.
These 27 equations can be solved to obtain the steady-state three-photon amplitude $EEE(x_1,x_2,x_3)$ inside the medium, \textit{i.e.}, within the range $\zero<x_{1,2,3}<\xo$.

To obtain the three-photon steady state, we numerically evolve the system over time until convergence at sufficiently large $t$.  The evolution is performed hierarchically on the different sections of the $(x_1,x_2,x_3)$ space in Fig.~\ref{fig:Dirac}b. We begin by solving for the steady-state wavefunction of a single polariton $\psi_i(x)$, with $i=\{E,P,S\}$, within the brown sections in Fig.~\ref{fig:Dirac}b. These solutions serve as (stationary) boundary conditions for the three light-blue sections \cite{Peyronel2012, firstenberg2013attractive,DroriScience2023}. 
We then compute the two-photon steady-state wavefunction $\psi_{ij}(x,x')$ within these sections, using the initial condition $\psi_{ij}=(x_1,x_2,t=0)=0$ and solving the nine equations of motion for two interacting polaritons (a reduced form of Eqs.~(\ref{eq:3_photon_equations}), \textit{e.g.}, Eq.~(S5) in Ref.~\cite{DroriScience2023}). Finally, these solutions are used as boundary conditions for the main (gray) section, 
\begin{gather}
\begin{aligned}
\psi_{ijk}(x_1=\zero,x_2,x_3,t)&=\psi_{i}(x_1=\zero)\psi_{jk}(x_2,x_3),\\
\psi_{ijk}(x_1,x_2=\zero,x_3,t)&=\psi_{j}(x_2=\zero)\psi_{ik}(x_1,x_3),\\
\psi_{ijk}(x_1,x_2,x_3=\zero,t)&=\psi_{k}(x_3=\zero)\psi_{ij}(x_1,x_2)\:,
\end{aligned}
\end{gather} 
and we solve Eqs.~(\ref{eq:3_photon_equations}) for the three-photon steady-state amplitude $\psi_{ijk}(x_1,x_2,x_3)$ with the initial condition $\psi(x_1,x_2,x_3,t=0)=0$. Examples of these results are shown in Fig.~\ref{fig:Chiral}.

\section{Phase diagram parameters}\label{app_phase_diagram_params}
\noindent  Here, we describe the parameters used for the phase diagram in Fig.~\ref{fig:Chiral}d. Following Eq.~(5) in the main text, we assume that the two-photon interaction is governed by a nearly-square potential $V^{(2)}(r)=U/(1+8r^6/r_{\rm b}^6)$, where $U$ represents the change in the refraction index due to the Rydberg blockade, {$U=\sqrt{2}\rho g^2/(\Delta c)$,} and the (negative) effective mass $m=-U/(2c)$ arises from single-photon dispersion.  The interaction strength is quantified by the parameter 
$\lambda= |U m|c r_{\rm b}^2$
and the interaction duration is quantified by $\varphi=U\len/\sqrt{2}$. 
We note that these expressions differ slightly from those in
Ref.~\cite{DroriScience2023} due to a different choice of rotated coordinates. Here we use $r=(x_1-x_2)/\sqrt{2}$ and 
$R=(x_1+x_2)/\sqrt{2}$, whereas in Ref.~\cite{DroriScience2023}, the definitions were $r=x_1-x_2$ and 
$R=(x_1+x_2)/2$.

\section{Propagation of two and three polaritons}\label{app_propagation}
\noindent  Our results for the two-polariton propagation in Fig.~\ref{fig:Dirac}(c,d), have been obtained by using a standard finite-difference scheme in the real space. We rewrite Eq.~\eqref{eq:two} in the center-of-mass frame, which is obtained from Eq.~\eqref{eq:K2} by replacing $\omegd$ in the last term by the interaction term $V_2$. Next, we replace the derivatives over $r$ and $R$ with the finite differences. We introduce a cutoff for $r=\pm 60r_b$, where we apply periodic boundary conditions, and propagate the solution from $R=0$ to $R={20 c/\omegd}$ with the initial conditions $ES_+=1$, $ES_-=0$. For the Schr\"odinger equation, we apply the same finite-difference scheme for Eq.~\eqref{eq:Shr}. 

Results for three interacting polaritons in Fig.~\ref{fig:Dirac}(e,f,g) have been obtained from  Eq.~\eqref{eq:three} where the dependence on $\upeta$ and $\upzeta$ is described in the reciprocal space. We outline the calculation technique below.

We solve Eq.~\eqref{eq:three} in the center of mass frame for
$\psi(\upeta,\upzeta,R)$ with the initial condition $\psi(\upeta,\upzeta,R=\zero)=[1,0,0]^{T}\:.$ 
The wavefunction is represented as 
\begin{equation}
    \psi(\upeta,\upzeta,R)=\sum\limits_{\bm b}\psi_{\bm b}(R)\e^{\rmi (b_\upeta\upeta+b_\upzeta\upzeta)}\:,
\end{equation}
where $\bm b$ are the reciprocal lattice vectors, defined as 
\begin{align}
&\bm b_{nm}=n\bm b_{1}+m\bm b_{2},\\ &\bm b_{1}=\frac{2\pi D}{S}\bm e_{1},~ \bm b_{2}=\frac{2\pi D}{S}\left(\frac{1}{2}\bm e_{1}+\frac{\sqrt{3}}{2}\bm e_{2}\right).
\end{align}
Here, $n,m$ are integers, $\bm e_{1,2}$ are Cartesian basis vectors, $D\gg r_\mathrm{b}$ is the lattice period, and $S=\sqrt{3}D^{2}/2$ is unit cell area in real space.
The numerical calculation in Fig.~\ref{fig:Dirac}(g-i) has been performed taking into account  
1069~vectors $\bm b$ and $D=120r_{\rm b}$.
Using this representation, Eq.~\eqref{eq:three} is transformed into a system of $3N$ linear equations, where $N$ is the chosen number of basis states,
\begin{equation}
\begin{split}
      \rmi\frac{\partial}{\partial R}\psi_{\bm b}&=
    \begin{pmatrix}
        0&b_{\upeta}&b_{\upzeta}\\
        b_{\upeta}&b_{\upzeta}/\sqrt{2}&b_{\upeta}/\sqrt{2}\\
        b_{\upzeta}&b_{\upeta}/\sqrt{2}&-b_{\upzeta}/\sqrt{2}\\
    \end{pmatrix} \psi_{\bm b}
    \\&{+}\frac{\sqrt{3}{\omegd}}{c}\psi_{\bm b}{-}\sum\limits_{\bm b'}\frac{\sqrt{3}V_{3,\bm b-\bm b'}}{c}\begin{pmatrix}
     1&0&0\\0&0&0\\0&0&0
     \end{pmatrix}
       \psi_{\bm b'}\:.
       \end{split}\label{eq:K3c}
\end{equation}
This expression is similar to Eq.~\eqref{eq:K3b} with the additional interaction  term.
Here,
\begin{equation}
V_{3,\bm b-\bm b'}=\frac1{S}\iint \rmd\upeta\rmd\upzeta \e^{-\rmi (b_\upeta\upeta+b_\upzeta\upzeta)}
V_{3}(\upeta,\upzeta)
\end{equation}
is the Fourier component of the interaction potential.
Following Eq.~\eqref{eq:m3}, the Schr\"odinger equation approximation is obtained by replacing Eq.~\eqref{eq:K3c} with the single-band model for a single scalar amplitude 
$\psi_{\bm b}$:
\begin{equation}
\begin{split}
      \rmi\frac{\partial}{\partial R}\psi_{\bm b}&=
      -\frac{c}{\sqrt{3}\omegd}(b_{\upeta}^2+b_{\upzeta}^2
      )\psi_{\bm b}
    \\&{+}\frac{\sqrt{3}{\omegd}}{c}\psi_{\bm b}{-}\sum\limits_{\bm b'}\frac{\sqrt{3}V_{3,\bm b-\bm b'}}{c}       \psi_{\bm b'}\:.
       \end{split}\label{eq:K3cs}
\end{equation}

We diagonalize Eq.~\eqref{eq:K3c} to find the eigenvectors $\psi_{\mu}^{(\bm b)}(R)\propto \e^{\rmi K_{\mu}R}$ with the corresponding eigenvalues $K_{\mu}$.
The solution is then expressed as 
\begin{equation}\label{eq:K3d}
    \psi(\upeta,\upzeta,R)=    \sum\limits_{\bm b,\mu}\e^{\rmi K_{\mu}R}\e^{\rmi(b_\upeta\upeta+b_\upzeta\upzeta)}\psi_{\bm b}^{(\mu)}\langle \mu|0\rangle\:,
\end{equation}
where $\langle \mu|0\rangle=[\psi_{0}^{(\mu)}]_{1}\delta_{\bm b,1}$ is the overlap of the eigenvector with the homogeneous initial solution.

\section{Photon dispersion in the full  model}\label{app_full_model}
Here we present the results of the numerical calculation of the  dispersion law
$K(\kappa_1\ldots \kappa_n)$ for the center-of-mass wave vector $K$.  We consider both $n=2$ and $n=3$ photons. As a starting point, we use either the full Hamiltonian of the single photon coupled to atoms 
\begin{equation}\label{eq:HS}
	H_{EPS}(k)= 	
	{-}\begin{pmatrix}
	-ck &  \sqrt{\rho}g & 0\\	
	 \sqrt{\rho}g   & \Delta+\rmi\Gamma &  \Omega \\
	0& \Omega & \delta+\rmi\gamma\\
		\end{pmatrix}
\end{equation}
(Eq. 1 in the main text) or the Hamiltonian with eliminated $p$ states 
\begin{equation}\label{eq:HPS}
	H_{ES}(k)= 	
		\begin{pmatrix}
	\omegd+ck &  \frac{\sqrt{\rho} g\Omega }{\Delta}\\	
\frac{\sqrt{\rho} g\Omega}{\Delta}   & \frac{\Omega^2}{\Delta}
		\end{pmatrix}\:,
\end{equation}\
where $k$ is the polariton wave vector.
%%%%%%%%%%%%%%%%%%%%%%%%%%%%%%%%%%%%%%%%%%%%%%%%%%%%%%%%
\begin{figure}[t!]
    \centering
       \includegraphics[width=0.5\linewidth]{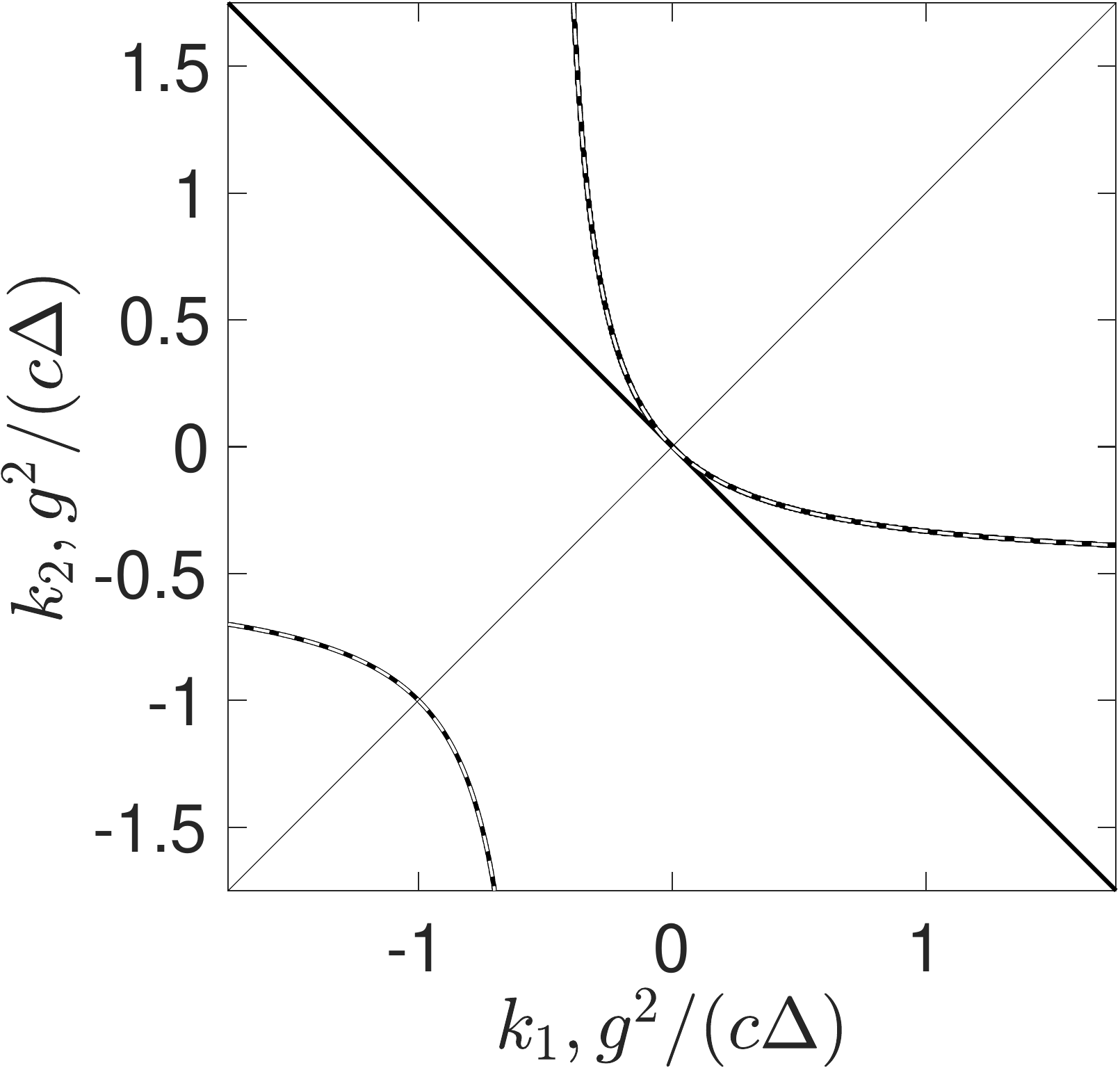}
    \caption{Two-photon dispersion law $\omega(k_1,k_2)=0$ under the stationary conditions.
    Dashed lines correspond to the dual-band model based on Eq.~(\ref{eq:two}), solid lines correspond to the 9-band model.
    Calculation has been performed for $\sqrt{\rho}g/\Delta=300$, $\Omega/\Delta=1/3$.}
    \label{fig:two-photon-dispersion}

\end{figure}
%%%%%%%%%%%%%%%%%%%%%%%%%%%%%%%%%%%%%%%%%%%%%%%%%%%%%%%%
Next, we construct the two-polariton and three-polariton Hamiltonians
\begin{align}
\mathcal{H}(k_1,k_2)
  &= H(k_1)\otimes I + I\otimes H(k_2), \\
\mathcal{H}(k_1,k_2,k_3)
  &= H(k_1)\otimes I\otimes I \notag \\
  &\quad + I\otimes H(k_2)\otimes I \notag \\
  &\quad + I\otimes I\otimes H(k_3).
\end{align}
Finally, we solve the Schr\"odinger equation 
$
\mathcal{H}\psi=\omega \psi 
$
to find  the dispersion law $\omega(k_1\ldots k_n)$. 
For example, for the two-photon model with eliminated $p$ states, we find 
\begin{align}
&\mathcal{H}(k_1,k_2)=\nonumber\\
&\resizebox{0.9\columnwidth}{!}{$
\begin{pmatrix}
ck_{1}+2\omegd+ck_{2}
& \dfrac{\sqrt{\rho}\, g \Omega}{\Delta}
& \dfrac{\sqrt{\rho}\, g \Omega}{\Delta}
& 0
\\
\dfrac{\sqrt{\rho}\, g \Omega}{\Delta}
& ck_{1}+\omegd+\dfrac{\Omega^{2}}{\Delta}
& 0
& \dfrac{\sqrt{\rho}\, g \Omega}{\Delta}
\\
\dfrac{\sqrt{\rho}\, g \Omega}{\Delta}
& 0
& ck_{2}+\omegd+\dfrac{\Omega^{2}}{\Delta}
& \dfrac{\sqrt{\rho}\, g \Omega}{\Delta}
\\
0
& \dfrac{\sqrt{\rho}\, g \Omega}{\Delta}
& \dfrac{\sqrt{\rho}\, g \Omega}{\Delta}
& \dfrac{2\Omega^{2}}{\Delta}
\end{pmatrix}
$}.
\end{align}
We consider the stationary regime, when $\omega=0$. Thus, we solve the equation $\omega(k_1\ldots k_n)=0$ numerically to find the center-of-mass wave vector $K=(k_1+\ldots +k_n)/\sqrt{n}$ as a function of the relative wave vectors given by
\begin{align*}
    &k=\frac{k_1-k_2}{\sqrt{2}}\quad (n=2) \quad\text{ and } \\
    &k_\upeta=\frac{k_1-k_2}{\sqrt{2}}, k_\upzeta=\frac{k_1+k_2-2k_3}{\sqrt{6}}\quad      (n=3)\:.
\end{align*}
A comparison of the results of the dual-band model and of the full 9-band model for two photons is presented in Fig.~\ref{fig:two-photon-dispersion}. The calculation demonstrates that the full model exhibits Dirac-like cones, which are satisfactorily reproduced by the model.

\begin{figure}[b!]
    \centering
    \includegraphics[width=0.99\linewidth]{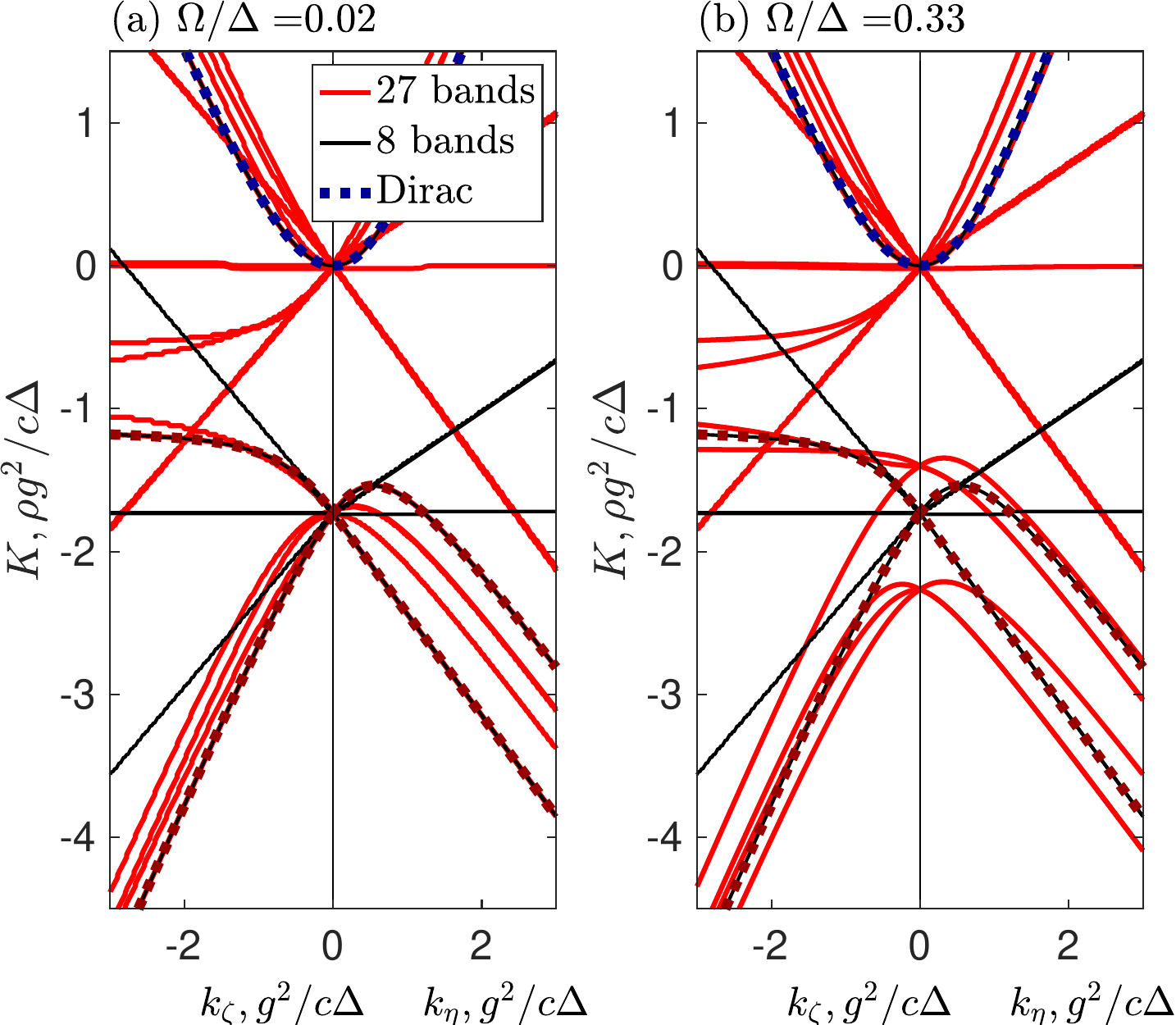}
    \caption{Three-photon multiband structure under the stationary conditions.
    Dotted lines correspond to the 3-band model based on Eqs.~(5), thin black lines to the 8-band model with eliminated $p$ states, thick red lines to the full 27-band model.
    Calculation has been performed for $\sqrt{\rho}g/\Delta=300$ and for two values of $\Omega/\Delta$ as indicated in the title.}
    \label{fig:three-photon-dispersion}
\end{figure}

Figure~\ref{fig:three-photon-dispersion} presents a comparison between the three-band model Eq.~(\ref{eq:three}), the 8-band model with eliminated $p$ states, and the full 27-band model. Both the 8-band model and the three-band model very well reproduce the $ESS_+$ band, which is relevant for experiments~\cite{DroriScience2023}. However,  the 27-band model features even more complicated dispersion with two split Dirac cones, instead of a single Dirac cone in 8-band and 3-band models. This splitting of the Dirac cones can be interpreted as an analogue of spin-orbit splitting in solid-state band structure, where the polaritons' degrees of freedom related to the $p$-state components play the role of a (pseudo)spin degree of freedom.  The two panels in Fig.~\ref{fig:three-photon-dispersion} are calculated for two different values of $\Omega/\Delta$, with panel (b) corresponding to the realistic value $\Omega/\Delta=1/3$~\cite{DroriScience2023}. Comparing these two panels, we see that the splitting between the Dirac cones increases with $\Omega/\Delta$.

\bibliography{bibliography}

\newpage\clearpage

\onecolumngrid 
\setcounter{figure}{0}
\renewcommand{\thefigure}{S\arabic{figure}}
\setcounter{equation}{0}
\renewcommand{\theequation}{S\arabic{equation}}

\end{document}